\documentclass[twocolumn,appendixfloats,tighten]{aastex6}
\usepackage{newtxtext,newtxmath}
\usepackage[T1]{fontenc}
\usepackage{ae,aecompl}

\usepackage{amsmath}	
\usepackage{amssymb}	
\usepackage{ctable}
\usepackage{url}
\usepackage{xspace}
\usepackage[normalem]{ulem}
\usepackage{hyperref}
\usepackage[all]{hypcap}
\usepackage{graphicx}
\usepackage{color}

\def\msun{{\rm\,M_\odot}}

\def\msun{{\rm\,M_\odot}}

\newcommand{\lya}{Ly$\alpha$ }

\newcommand{\be}{\begin{equation}}
\newcommand{\ee}{\end{equation}}

\def\h2{${\rm\,H_2}$}

\def\lya{Ly-$\alpha~$}

\def\gsim{ \lower .75ex \hbox{$\sim$} \llap{\raise .27ex \hbox{$>$}} }
\def\lsim{ \lower .75ex \hbox{$\sim$} \llap{\raise .27ex \hbox{$<$}} }

\usepackage{color}

\begin{document}

\title{ A limit on the warm dark matter particle mass from the redshifted 21 cm absorption line}

\author{Mohammadtaher Safarzadeh\altaffilmark{1}, Evan Scannapieco\altaffilmark{1},  Arif Babul\altaffilmark{2}}

\altaffiltext{1}{ School of Earth and Space Exploration, Arizona State University, \href{mailto:mts@asu.edu}{mts@asu.edu}}
\altaffiltext{2}{ Department of Physics and Astronomy, University of Victoria, Victoria, BC, V8P 5C2, Canada}

\begin{abstract}
The recent EDGES collaboration detection of an absorption signal at a central frequency of $\nu = 78 \pm 1$ MHz points to the presence of a significant Lyman-$\alpha$ background by a redshift of $z=18$. The timing of this signal constrains the dark matter particle mass ($m_\chi$) in the warm dark matter (WDM) cosmological model. WDM delays the formation of small-scale structures, and  therefore a stringent lower limit can be placed on $m_\chi,$ based on the presence of  a sufficiently strong \lya background due to star formation at $z=18$, 
Our results show that the coupling the spin temperature to the gas through \lya pumping requires a minimum mass of  $m_\chi>3$ keV if atomic cooling halos dominate the star formation rate at $z=18,$ and $m_\chi>2$ keV if \h2 cooling halos also form stars efficiently at this redshift. These  limits match or exceed the most stringent limits cited to date in the literature, even in the face of the many uncertainties regarding star-formation at high redshift.

\end{abstract}

\section{Introduction}

Recently the EDGES collaboration \citep[Experiment to Detect the Global Epoch of reionization Signature; ][]{Bowman:2010fk} detected a global absorption signal at a central frequency of $\nu =78 \pm 1$ MHz,  with a brightness temperature $T = -500^{+200}_{-500}$ mK and a signal-to-noise ratio of $\approx 37$ \citep{Bowman:2018bs}. 
This points to the indirect coupling of the spin temperature ($T_S$) to the gas temperature ($T_K$) through the Wouthuysen-Field (WF) effect \citep{Wouthuysen:1952hd,Field:1957fa} by a background of \lya photons at a redshift of $z = 17.2,$ which corresponds to $\approx$ 180 million years after the Big Bang.

The spin temperature of HI is related to the gas ($T_K$), cosmic microwave background ($T_\gamma$), and \lya background ($T_\alpha$) temperatures  by
\be
T_S^{-1}=\frac{T_\gamma^{-1} + x_c T_K^{-1} + x_\alpha T_{\alpha}^{-1}}{1+x_\alpha+x_c},
\ee
where $x_c$ is the collisional coupling constant and $x_\alpha$ is the coupling coefficient to \lya photons \citep{Chen:2004ex,Hirata:2006kr} given by
\be
x_\alpha=\frac{8 \pi \lambda_{\rm Ly \alpha}^2 \gamma T_*}{9A_{10} T_\gamma} S_\alpha J_\alpha = 1.81\times10^{11} \frac{J_\alpha S_\alpha }{1+z},
\ee
where $\gamma$ = 50 MHz is the half width at half-maximum  of the \lya resonance, 
 $T_*=h\nu_{10}/k_B$ = 68.2 mK with $\nu_{10} = 1.42$ GHz, $A_{10}=2.87\times10^{-15} s^{-1}$ is the spontaneous emission coefficient of the 21 cm line, $J_\alpha$ is the background \lya intensity (in units of $\rm cm^{-2} s^{-1} Hz^{-1} sr^{-1}$), and $S_\alpha $ is a factor that accounts for spectral distortions, which is less than 1 when the spin and gas temperature are similar \citep{Hirata:2006kr}.  %

As $x_\alpha$ increases, the spin temperature shifts from the microwave background temperature to $T_\alpha,$ providing information about the temperature of the high-redshift intergalactic medium.  Interestingly, the best fit value of this temperature is lower than expected, potentially suggesting a subcomponent of dark matter particles with masses in the MeV range and charges a fraction of that of the electron \citep{Dvorkin:2014if,Munoz:2018vz,Barkana:2018dq,Barkana:2018tj}. 

At the same time, warm dark matter (WMD) candidates with masses in the keV range have been proposed to account for several tensions between observations and theoretical predictions.  In particular, (i) cold dark matter (CDM) cosmological simulations predict cuspy halos profiles while  observations point to more core-like centers \citep{Bullock:2001hp,Gentile:2004ba}; 
(ii) CDM simulations predict larger stellar velocity dispersions than observed in  Milky Way's satellite galaxies \citep{BoylanKolchin:2012id}; and 
(iii) the number of subhalos predicted in the CDM simulations far exceeds the observed number of luminous  Milky Way satellites \citep{Klypin:1999ej,Moore:1999ja}.

While it remains a matter of debate whether the resolution of these tensions can be achieved by improving models of baryonic physics,  WDM particles could also address these issues by suppressing the formation of small-scale structure.  These particles, such as sterile neutrinos \citep{Dodelson:1994ff,Abazajian:2001dx} and gravitinos \citep{Gorbunov:2008fr},  remain relativistic for a longer time in the universe.  This leads to non-negligible velocity dispersions which causes them to free stream out of small scale perturbations, 
attenuating the matter power spectrum above a characteristic comoving wave number \citep{Bode:2001ie,Viel:2005gk} of
\be
 k_{\rm FS} =15.6 \frac{h}{\rm Mpc} \left( \frac{m_\chi}{1 \rm keV} \right)^{4/3} \left(\frac{0.12}{\Omega_{\rm DM}h^2} \right)^{1/3},
\ee
which leads to less pronounced cusps and fewer and smaller-mass Milky Way satellites.

Various approaches have been adopted to provide lower bounds on $m_\chi$, which in turn place lower limits on  $k_{\rm FS}.$
Based on the abundance of $z=6$ galaxies in the Hubble Frontier Fields, \citet{Menci:2016bc} arrive at $m_\chi>2.4$ keV ($2\sigma$) and \citet{Corasaniti:2017fw} arrive at  $m_\chi>1.5$ keV ($2\sigma$) based on the galaxy luminosity function at $z \approx 6-8.$ Similarly, \citet{Pacucci:2013dd} conclude that lensing surveys such as CLASH provide  $m_\chi>0.9$ keV ($2\sigma$) lower bounds.
The current best lower limit of  $m_\chi>3.3$ keV ($2\sigma$) is based on the high redshift \lya forest data \citep{Viel:2013he}. 

The detection of the redshifted 21-cm background is also able to provide  a limit on  $m_\chi,$ based on the simple fact that enough small-scale structures must collapse at high redshift to lead to a  \lya background strong enough to couple $T_S$ to $T_\alpha$ at the high end of the absorption trough, just beyond $z=18.$ The greater the suppression of small dark-matter halos, the more difficult it is for \lya to couple the spin and the gas temperatures by the WF effect, and thus by calculating the maximum \lya intensity allowed as a function the rate of dark matter collapse, we are able derive a minimum allowed value for the WDM particle. Although the idea of inferring the WDM particle mass from redshifted 21-cm absorption line has been proposed before \citep{Yoshida:2003cl,Sitwell:2014ez}, we have applied the idea to the EDGES signal in this paper.

The structure of this work is as follows.
In \S2 we show how the high-redshift halo mass function depends on $m_\chi$, and relate this mass function to the \lya background. 
In \S3 we show how the delayed formation of halos would place lower limits on $m_\chi,$ depending on whether atomic or \h2 cooling
halos dominate the star formation at $z=18$. 

\section{method}

\subsection{The Halo Mass Function}

\begin{figure}
\resizebox{3.5in}{!}{\includegraphics{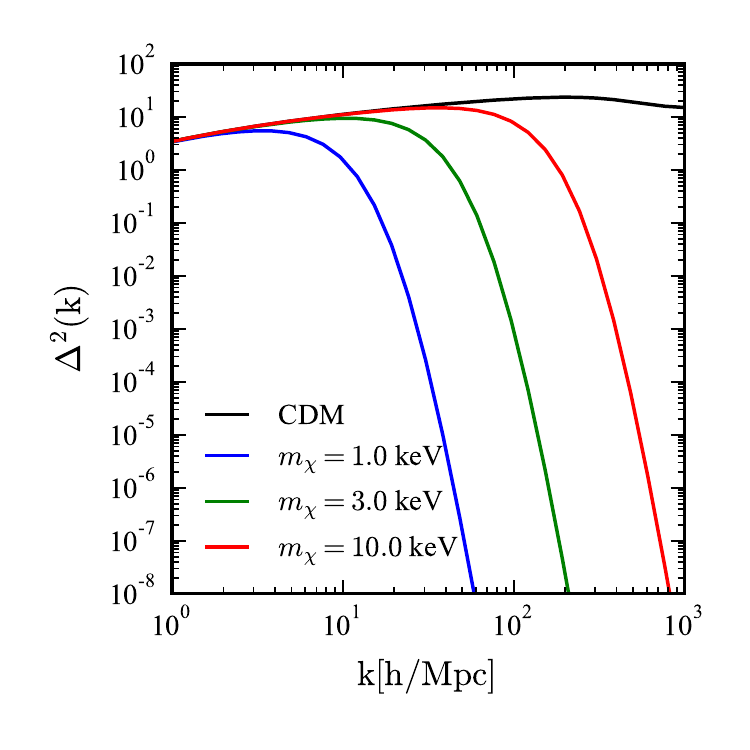}}
\caption{The normalized power spectrum $\Delta^2(k)=k^3P(k)/(2\pi^2)$ for different cosmological models. 
The black line corresponds to the CDM model, while the blue, green, red and red lines correspond to the m=1, 3, and 10 keV WDM models respectively.}
\label{f.deltak}
\end{figure}

Throughout this paper, we adopt the Planck 2015 cosmological parameters \citep{Collaboration:2016bk} where $\Omega_M=0.308$, $\Omega_\Lambda=0.692$, $\Omega_b=0.048$ are total matter, vacuum, and baryonic densities, in units of
the critical density $\rho_c$, $h=0.678$ is the Hubble constant in units of  100 km/s/Mpc, $\sigma_8= 0.82$ is the variance of linear fluctuations on the
8 $h^{-1}$ Mpc scale, $n_s=0.968$ is the tilt of the primordial power spectrum, and $ Y_{\rm He}=0.24$ is the primordial helium fraction.
We compute the power spectrum for CDM and modify it according to the following formula to be appropriate for different WDM scenarios
with different particle masses
\be
P_{\chi}(k) = T^2(k) P_{CDM} (k),
\ee 
where $P(k)$ denotes the power spectrum as a function of comoving wavenumber $k$. The power spectrum is taken from CAMB \citep{Lewis:2000ju},
and we adopt the fitting formula for $T(k)$ given by \citet{Bode:2001ie}:
\be
T(k)= \left[ 1+(\alpha k )^{2\mu} \right]^{-5/\mu},
\ee
where $\mu=1.12$ and the parameter $\alpha$ determines the cutoff position in the power spectrum as 
    \begin{multline}
      \alpha = 
      \frac{0.05}{h\rm{Mpc}^{-1}}
      \left(\frac{m_{\rm{x}}}{1\rm{keV}}\right)^{-1.15}
      \left(\frac{\Omega_{\rm{WDM}}}{0.4}\right)^{0.15} \\ 
      \times\left(\frac{h}{0.65}\right)^{1.3}
      \left(\frac{g_{\rm{WDM}}}{1.5}\right)^{-0.29},
    \end{multline}
where $\Omega_{\rm{WDM}}$ is the WDM contribution  to the density parameter, and we set the number of degrees of freedom to $g_{\rm{WDM}}=1.5$.  
Figure \ref{f.deltak} shows the power spectrum for the CDM model as compared to WDM model for $m_\chi=$1, 3, and 10 keV.
  
\begin{figure*}
{\vspace*{1cm}
{\includegraphics[width=1\linewidth]{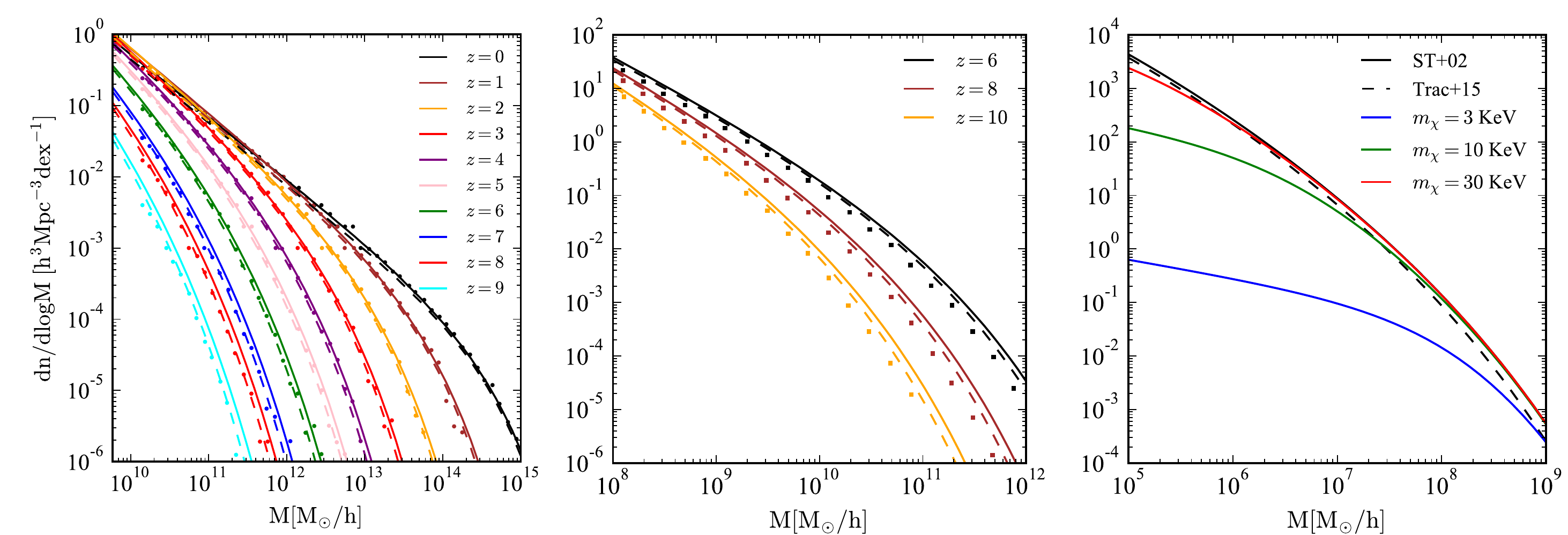}}}
\caption{{\em Left panel:} The redshift evolution of our HMFs (lines) compared to the Bolshoi-Planck N-body simulations (points) out to $z=9$ \citep{RodriguezPuebla:2016kr}. 
Solid lines correspond to HMFs parametrized as per ST02 while dashed lines show the HMFs based on \citet{Trac:2015by} fits.
The exact redshift outputs from Bolshoi simulation are: $z=0,1.016,2.04,3.032,4.067,4.989,6.058,7.351,8.122,9.404$. 
{\em Middle panel:} Comparing the ST02 (solid lines) and Trac15 (dashed lines) HMFs to \citet{Trac:2015by} data points which resolves smaller halos compared to Bolshoi. {\em Right panel:} The HMF at $z=18$. 
The black solid and dashed lines corresponds to the CDM model based on ST02 and Trac15 HMF parametrization, while the blue, green and the red lines correspond to WDM models with $m_\chi=$3, 10, and 30 keV respectively.}
\label{f.dndlnM}
\end{figure*}

We compute the halo mass function (HMF) as  \citep{Press:1974jb}:
\begin{equation}\label{dndlnM}
\frac{dn}{d\ln M}=\frac{\bar{\rho}}{M}f(\nu)\frac{d\nu}{d\ln M},
\end{equation} 
where $n$ is the number density of haloes, $M$ the halo mass, $\nu$ the peak-height of perturbations,  $\bar\rho$ the average density of the universe, and the 
first crossing distribution $f(\nu)$ \citep{Bond:1991cc} is obtained 
from the ellipsoidal collapse model leads as 
\begin{equation}\label{nufnu}
\nu f(\nu) = A \sqrt{\frac{a\nu}{2\pi}}\left[1+(a\nu)^{-p}\right]e^{-a\nu/2},
\end{equation}
with $A=0.322$, $p=0.3$, and $a=0.75$ \citep[][hereafter ST02]{Sheth:2002fn}. 
Here, the peak height $\nu$ is defined as $\nu \equiv \delta_{c,0}^2 {\sigma_\chi(R,z)}^{-2},$ with $\delta_{c,0}=1.686$. The variance  is  $\sigma^2_\chi(M,z)= \sigma^2_\chi(M,0) D(z)^2,$ with
\be 
\sigma^2_\chi(M,0)= \sigma^2_\chi(R,0)= \int_0^{\infty}\frac{dk}{2 \pi^2} \,k^2 P_\chi(k) w^2(kR)\ ,\label{eqsigM}
\ee 
where $M = 4 \pi R^3 \Omega_M \rho_c/3$, $w(kR)\equiv 3 j_1(kR)/(kR),$ with $j_1(x)\equiv(\sin x-x\cos x)/x^2,$ and $D(z)$ is the linear growth factor 
\be
D(z)\equiv\frac{H(z)}{H(0)} \int_z^{\infty} \frac{dz^\prime (1+z^\prime)}{H^3(z^\prime)} \Bigg [\int_0^{\infty} \frac{dz^\prime (1+z^\prime)}{H^3(z^\prime)}\Bigg]^{-1}.
\ee
While the adoption of a spherical top-hat window filter is not perfect for a truncated power spectrum \citep{Benson:2013ir}, it is a conservative choice, in that it gives a weaker lower bound on $m_\chi$ than
other window functions in the literature. Figure \ref{f.dndlnM} shows the redshift evolution of the HMF compared to the Bolshoi-Planck N-body simulations from $z=0$ to 9 \citep{RodriguezPuebla:2016kr} and the    \citet{Trac:2015by} (Trac15) simulations at$z=$6, 8, and 10.   Our formula provides a good match to all these results, and the right panel of Figure \ref{f.dndlnM} shows how this function changes for different values of $m_\chi$ at $z=18$.

\vspace*{2cm}

\subsection{The Emissivity of Sources}
The comoving \lya  photon intensity at redshift $z$ is given by \citep{Chen:2004ex,Hirata:2006kr}:
\be
J_\alpha(z)=\frac{1}{4 \pi} (1+z)^2\sum\limits_{n=2}^{\infty} P_{n{\rm p}} \int_z^{z_{\rm max}} \frac{c}{H(z^\prime)} \epsilon(\nu^\prime,z^\prime) dz^\prime,
\ee
where 
\be
z_{\rm max}=\frac{1-(n+1)^{-2}}{1-n^{-2}}(1+z) -1,
\ee
 is the horizon beyond which photons redshift into another resonance line,
 $P_{n{\rm p}}$ is the probability of producing a \lya photon following the excitation of HI to the $n$p configuration \citep[as given by Table 1 of][]{Hirata:2006kr},
and
\be
\epsilon(\nu^\prime,z)=\epsilon_b(\nu_{\alpha}) \times \left(\frac{\nu^\prime}{\nu_\alpha} \right)^{\alpha_s-1} \tau_* \frac{d}{dt} \int_{M_{\rm min}}^{\infty} n(M,t) f_*(M) M dM
\ee
is the comoving photon emissivity (defined as the number of photons emitted per unit comoving volume, time, and frequency) at a redshift $z^\prime$ and frequency $\nu^{\prime},$
where $n(M, t)$ is the comoving number density of haloes  per unit mass, $f_*(M)$ is the fraction of the mass that is turned into stars as a function of halo mass, and 
\be
\epsilon_b(\nu_{\alpha})=\frac{L_\alpha}{h\nu_\alpha}.
\ee
Here $L_\alpha=2.7\times10^{21} \rm erg s^{-1} Hz^{-1 }\msun^{-1}$ for Population III stars,  $\tau_*=2$ Myr is the main sequence lifetime of very massive stars \citet{Bromm:2001ih}, and $\alpha_s=1.29$ \citep{Barkana:2005bl}. \citet{Scannapieco:2003io} arrive at a slightly larger value of $L_\alpha=3.37\times10^{21}\rm erg s^{-1} Hz^{-1 }\msun^{-1}$  with a similar lifetime for Population III stars in the mass range of 50-500$\msun$ with a top-heavy Salpeter IMF. This is by a factor of $\approx 3$ larger than the \lya yields from Population III stars in the 1-500 $\msun$ mass range. To be on the conservative side, we adopt the largest possible \lya yield of $L_\alpha=3.37\times10^{21}\rm erg s^{-1} Hz^{-1 }\msun^{-1}$ in this paper.

The lower limit of the integral is set based on whether we assume atomic cooling halos dominate the SFR or the $\rm H_2$ cooling halos. 
In the case of atomic cooling halos, the limit is $ M_{\rm min}=4\times 10^7\msun$ at $z=18$ where it corresponds to halos with virial temperature of $\rm T_{vir}>10^4$K. 

The minimum halo mass in order for the gas to cool and condense to form stars due to both atomic and \h2 cooling based on AMR simulations \citep{Machacek:2001fq,Wise:2007ds} is:
\be\label{m_crit}
M_{\rm crit}=2.5\times10^5 + 1.7 \times 10^6 (F_{LW}/10^{-21})^{0.47} \msun,
\ee
where $F_{LW}$ is Lyman-Werner intensity integrated over solid angle in units of ergs s$^{-1}$ cm$^{-2}$ Hz$^{-1}$.
To be conservative, we consider all halos with $M>2.5\times10^5\msun$, ignoring the suppression of star formation due to $F_{LW}$.

Finally, $f_*$ is modeled as $f_*= \epsilon _* \times f_b$ where $f_b=\Omega_m/\Omega_M$. 
For atomic cooling halos we adopt a conservative value for star formation efficiency (SFE) of $\epsilon_*=3\%$. This is a generous choice as it leads to an SFR density of 0.033 $\rm \msun/year/cMpc^3$ at $z=8$ which is a factor of 3 above the upper bounds of the observed SFR density of $\log \dot{\rho}_*=-2.08\pm{0.07} M_{\odot} \, \rm{yr}^{-1} \, \rm{cMpc}^{-3}$ at $z=8$ \citep{Madau:2014gt,Bouwens2015}. 
The observed SFR density at $z=8$ is obtained by integrating the observations down to a limiting magnitude of $M_{\rm AB}=-17$ and correcting for dust obscuration 
 \citep[see Table 7 of][]{Bouwens2015}. We choose a value 3 above the observed upper limit to ensure we take into account the faint population of galaxies that are not observed directly. 
Our high SFR density is consistent with the maximum extrapolated upper limits on SFR density based on SFE models of \citet{SF2016}.
For $\rm H_2$ cooling halos we similarly assume $\epsilon_*=3\%,$ even though this would require almost all the cold gas found in
AMR simulations of early star formation to be converted into stars \citep{Machacek:2001fq,Wise:2007ds}. 

\section{Results}
\begin{figure*}
{\includegraphics[width=1\linewidth]{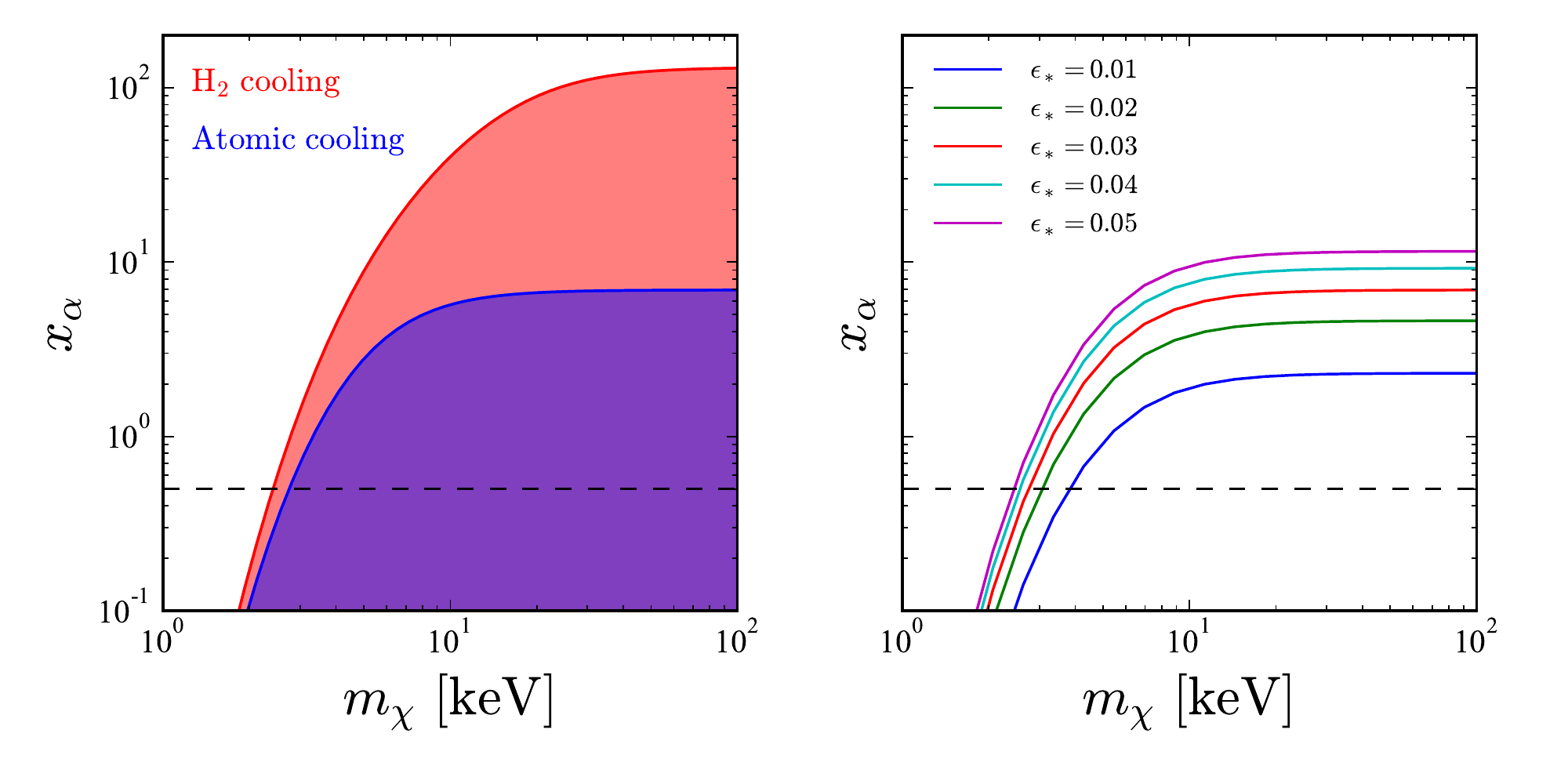}}
\caption{\lya coupling coefficient in the WDM model as a function of $m_\chi$.  {\em Left panel}: The solid red line shows the $x_\alpha$ 
when \h2 and atomic cooling halos both contribute to the SFR density at $z=18$ while the blue line shows the predicted $x_\alpha$ when only considering the atomic cooling halos. 
We adopt $\epsilon=3\%$ with minimum halo mass for star formation to be $2.5\times10^5\msun$ for \h2 cooling halos based on 
numerical simulations of early structure formation \citep{Machacek:2001fq,Wise:2007ds}. For atomic cooling halos we adopt $\epsilon_*=3\%$ in order to be conservative for atomic cooling halos. The red (blue) shaded regions refer to the predicted range of $x_\alpha$ in the two models. The black dashed line is $x_\alpha=0.5$. {\em Right panel}: The impact of changing the SFE for atomic cooling halos from 1\% to 5\%. A lower limit of 4 keV can be gotten with a less conservative approach and assigning a low SFE of $\epsilon_*=1\%$ to the halos.}
\label{f.xalpha}
\end{figure*}

Left panel of Figure \ref{f.xalpha} shows the \lya coupling coefficient based on the estimated flux of \lya photons in WDM model as a function of $m_\chi$ for two cases: (i)  only atomic cooling halos contributing to  star formation at $z=18$, and (ii) both \h2 cooling and atomic cooling halos contributing to $z=18$ star formation.  Here we conservatively adopt $S_\alpha=1$.  In the right panel we study the impact of changing the SFE for the atomic cooling halos. Requiring coupling of gas to spin temperature to have $x_\alpha>0.5$ through \lya pumping by Population III stars translates into a lower bound of $m_\chi>2~(3)$ keV if  \h2 (atomic) cooling halos dominate the SFR density at $z=18$ respectively.  

Fitting for the EDGES signal and the ultraviolet luminosity function (UVLF) of galaxies out to $z\approx10$, \citet{Mirocha:2018us} predicts an SFR density of $\approx10^{-3}$ and $2 \times10^{-2}$ $\rm \msun yr^{-1} cMpc^{-3}$ at $z\approx 18$ and $z\approx10$ respectively when considering only atomic cooling halos (see their Figure 4 right panel). 
Adopting a SFE of 3\%, we predict an SFR density of $\approx2.1\times10^{-3}$ and $2.6\times 10^{-2}$ at the same redshifts without including the \h2 cooling halos. 

The various different sets of cosmological parameters reported by Planck team show $< 2\%$ fractional differences in  $\Omega_M^{0.3} \times \sigma_8$,  a proxy for structure formation power \citep{Collaboration:2016bka,Collaboration:2016if}. We would have 4\% less mass in collapsed halos had we adopted cosmological parameters from counting clusters \citep{Collaboration:2016kg,Salvati:2017cb}., and about 3\%  more collapsed mass had we adopted WMAP9 cosmological parameters \citep{Hinshaw:2012dd}. We have verified that our results are not sensitive to such variations. 

Of the various cosmological parameters, the one that impacts our calculations the most is $\sigma_8$. Over the past decade, the reported values of $\sigma_8$,
from studies of galaxy clusters to the CMB, vary from 0.75 to 0.88.  In order to ascertain the sensitivity of our lower bound on $m_\chi$ to the adopted value of $\sigma_8$, we repeat the calculation for the two endpoint values of $\sigma_8$, while keeping the other cosmological parameters fixed to their fiducial values {\it and} requiring the same 3 times the upper limits of the observed SFR density at $z=8$. Only values of $\sigma_8$ higher than our fiducial value of 0.82 affect the outcome, with the lower bound dropping to 2 keV (for $\sigma_8$ =0.88) in the case of only atomic cooling halos. However, values of $\sigma_8>$ 0.88 are dated and nearly all recent determinations of $\sigma_8$, whether based on CMB or clusters, result in $\sigma_8 \lsim~0.82$.

Additionally, our calculations are based on the assumption that the SFE is the same for all the halos. As an alternative, \citet{Mirocha:2018us} parameterize the SFE as a function of the halo mass, 
such that it puts a greater emphasis on the more massive halos to provide the \lya photons. The drop in SFE towards lower halo masses, which is consistent with observations at low redshifts \citep{Behroozi:2013fga}, would boost our limit on $m_x$ if one maintains the same normalization of the SFE in the observed range. This is because the number density of halos in the atomic cooling range at $z=18$ that could provide the \lya budget is effectively limited to $10^8-10^9\msun/h$ as more massive halos are rarely present at such early epochs.  Assigning a lower SFE to such halos would make their presence more crucial to provide the required \lya budget, which would lead to a more stringent limit on $m_\chi$ as less suppression is allowed. The bounds on $m_\chi$ become weaker if either small halos are much better at forming stars than the ones observed or
the SFE increases such that it compensates the exponential drop in number density of atomic cooling halos at $z=18$. 
Then we can have more suppression and still reproduce the EDGES results.  
But no observational or theoretical studies support such trends.

Also, while we have assumed that the SFE does not change with redshift, we cannot rule the possibility that it begins increasing beyond $z=8$ in a way that is not seen in either numerical simulations or in lower redshift observations \citep[see][and references therein]{SF2016}. Similarly, our choice of $S_\alpha=1$ is motivated by the fact that lower values of $S_\alpha$ would push the lower bound on $m_x$ to larger values.

We note that we have not modeled the shape of the EDGES signal in this paper, which implies that we have assumed the timing of the signal and its shape could be studied separately. If instead these two are related, we would not be able to make the arguments in the paper without properly modeling the absorption signal.

These lower limits match or exceed the most stringent limits achieved so far in the literature \citep{Viel:2013he}, even in the face of the many uncertainties regarding star-formation at high redshift.  As observations and models of high-redshift galaxies continue to improve, along with detections of redshift 21 cm absorption line, comparisons of the type described here will continue to provide new insight and constraints on warm dark matter and its role in the history of structure formation.

\acknowledgements
We are thankful to the referee for their constructive comments. We are also thankful to Jordan Mirocha, Andrew Benson, Marc Kamionkowski, Aldo Rodriguez and Joel Primack for helpful discussions. 
This work was supported by the National Science Foundation under grant AST17-15876, by NASA under theory grant NNX15AK82G, 
and by NSERC (Canada) through the Discovery Grant program.

\bibliographystyle{apj}

\end{document}